\documentclass[journal]{IEEEtran}
\usepackage{xcolor,soul,framed} 
\colorlet{shadecolor}{yellow}

\usepackage[pdftex]{graphicx}
\graphicspath{{../pdf/}{../jpeg/}}
\DeclareGraphicsExtensions{.pdf,.jpeg,.png}

\usepackage[cmex10]{amsmath}
\usepackage{xcolor}
\usepackage{array}
\usepackage{eqparbox}
\usepackage{url}
\usepackage{orcidlink}
\usepackage{amssymb} 

\hypersetup{colorlinks=true, urlcolor=blue, linkcolor=blue, citecolor=blue,plainpages=false} 

\makeatletter
\renewcommand\subsubsection{\@startsection{subsubsection}{3}{\z@}%
   {1.5ex plus .5ex minus .2ex}%
   {0.5ex plus .2ex}%
   {\normalfont\normalsize\bfseries}}
\makeatother

\newcommand{\meanN}{\bar{N}}   

\begin{document}
\bstctlcite{IEEEexample:BSTcontrol}
    \title{Hypothesis-Based Particle Detection for Accurate Nanoparticle Counting and Digital Diagnostics}
  \author{Neil~H.~Kim\orcidlink{0000-0001-9966-4998},
  Xiao-Liu Chu\orcidlink{0000-0002-7476-2097}, 
  Joseph B. DeGrandchamp\orcidlink{0000-0003-2194-5495},
      Matthew~R. Foreman\orcidlink{0000-0001-5864-9636}~\
  \thanks{This work was funded by the Institute for Digital Molecular Analytics and Science (IDMxS) under the Singapore Ministry of Education Research Centres of Excellence scheme (EDUN C-33-18-279-V12) and Ministry of Education Academic Research Fund (Tier 1) Grant RG137/24. The authors are with the Institute for Digital Molecular Analytics and Science, Nanyang Technological University, 59 Nanyang Drive, Singapore 636921, Singapore. M.R. Foreman is also with the School of Electrical and Electronic Engineering, Nanyang Technological University, 50 Nanyang Avenue, Singapore 639798. email: matthew.foreman@ntu.edu.sg}
  }  
\markboth{
}{Kim \MakeLowercase{\textit{et al.}}: TTTTTTT}
\maketitle

\begin{abstract}

Digital assays represent a shift from traditional diagnostics and enable the precise detection of low-abundance analytes, critical for early disease diagnosis and personalized medicine, through discrete counting of biomolecular reporters. Within this paradigm, we present a particle counting algorithm for nanoparticle based imaging assays, formulated as a multiple-hypothesis statistical test under an explicit image-formation model and evaluated using a penalized likelihood rule. In contrast to thresholding or machine learning methods, this approach requires no training data or empirical parameter tuning, and its outputs remain interpretable through direct links to imaging physics and statistical decision theory. 
 
Through numerical simulations we demonstrate robust count accuracy across weak signals, variable backgrounds, magnification changes and moderate PSF mismatch. Particle resolvability tests further reveal characteristic error modes, including under-counting at very small separations and localized over-counting near the resolution limit.  Practically, we also confirm the algorithm's utility, through application to experimental dark-field images comprising a nanoparticle-based assay for detection of DNA biomarkers derived from SARS-CoV-2. Statistically significant differences in particle count distributions are observed between control and positive samples. Full count statistics obtained further exhibit consistent over-dispersion, and provide insight into non-specific and target-induced particle aggregation. These results establish our method as a reliable framework for nanoparticle-based detection assays in digital molecular diagnostics.

\end{abstract}


\IEEEpeerreviewmaketitle

\section{Introduction}
Next generation biological assays require very low limits of detection; that is the ability to detect a few copies of specific biomarkers (e.g. DNA strands) or pathogens (e.g. virus particles) in large sample volumes, to enable confident, accurate and early diagnoses. Digital assays, are an emerging cheap and convenient approach that achieves the requisite sensitivity by partitioning the sample and \emph{counting} discrete positive molecular detection events among partitions (which maps directly to analyte abundance), rather than conventional bulk assays which fit continuous readouts to calibration curves. This fundamental digital counting principle can enable measurements of extremely low analyte concentrations (attomolar to picomolar), whilst also simultaneously reducing dependence on absolute calibration standards, which can suffer from nonlinearity, temporal drift, and sensitivity to matrix and batch effects \cite{duffy2023digital,basu2017digital,walt2013perspective}. In imaging-based digital assays, nanoparticles or fluorophores can act as discrete reporters of analyte presence. Whilst myriad protocols have been reported \cite{PARK2024135670, Maley2020,  liu2022single,monroe2013single,aguirre2018csr,  Ferrari2023, McAffee2025.08.14.670347},
each fundamentally requires counting the number of spot-like images of each reporter to provide direct detection or calibration-free quantification of the target. Accurate detection and counting of such localized features in  images, however, remain challenging, especially in high-throughput digital diagnostics where low signal-to-noise ratios, heterogeneous backgrounds, and the need to resolve dense clusters of localized signals complicate reliable quantification. Distinct imaging modalities used in digital assays (such as dark-field, interferometric scattering or fluorescence imaging) can furthermore introduce specific intricacies related to background inhomogeneity, unique noise profiles, and variability in the point spread function (PSF) \cite{daaboul2014spiris, SUN2018513, Lew:11}.  

 Localization, counting and tracking of spot like objects is a common task, not only in diagnostic assays, but also in fluorescence and superresolution microscopy, nanometrology and astronomy. Spot detection methods have therefore been developed and evaluated in earnest \cite{Cheezum2001, Chenouard2014, AnalChemReview2014, Mabaso_Withey_Twala_2018}. As we discuss further in Section~\ref{sec:lit_rev}, conventional approaches  often rely on filter and threshold based processing or, more recently, machine learning (ML). Filtering and thresholding based approaches however inherently introduce some degree of arbitrariness that reduces reliability \cite{Mabaso_Withey_Twala_2018, Smal2010,  small2014accurate}. ML methods, while powerful for biomedical image segmentation \cite{ronneberger2015unet}, object recognition \cite{he2017maskrcnn} and particle detection \cite{boyle2023iscatmaskrcnn}, typically require extensive annotated datasets and careful domain-specific tuning, whilst also often producing outputs that are difficult to interpret in terms of underlying image physics. 
 
  To address these challenges, we build on the successes of recent  statistical based approaches \cite{mortensen2010optimized,smith2015mboc,Tang2016} and introduce a physics-grounded alternative that treats particle counting as a statistical decision problem under an explicit image-formation model, in which each hypothesis corresponds to a distinct particle count. Statistical decision theory is subsequently used to select the hypothesis that best explains the observed image \cite{kay1998detection} with an information-criterion style complexity correction. This eliminates the need for empirical thresholds or training data, directly links inference to measurable noise statistics and optical parameters, and yields outputs that remain interpretable in terms of physical imaging properties.
In Section~\ref{sec:methods} we describe the theoretical framework of our proposed algorithm and its implementation. We further perform a comprehensive evaluation using simulated images (Sections~\ref{sec:synthetic}) where ground truth is precisely defined, hence enabling systematic characterization across a wide range of imaging parameters. From such benchmarking of detection performance we derive practical strategies for optimizing acquisition settings (Sections~\ref{sec:accuracy_results} and ~\ref{sec:resolvability_results}). To complement these results and verify practical utility of our algorithm, we perform an experimental nanoparticle imaging assay closely approximating SARS-CoV-2 genomic detection using stable DNA mimics (Section~\ref{sec:exp_results}). Clear statistically significant differences are seen in particle count distributions providing robust analyte detection.
Together, these results establish the method as a reliable tool for nanoparticle imaging diagnostic assays and a principled framework for digital molecular detection.

\section{Related Work}\label{sec:lit_rev}
Current approaches for spot detection can broadly be classified into more traditional computational filter based approaches, ML, and statistically informed methods. The former typically consists of denoising, filtering and detection steps. Initial image preprocessing aims to enhance the spots relative to the background and noise. This often starts with basic corrections, such as mean subtraction and flat field correction, to account for camera-specific artifacts like dark currents \cite{Janesick1987, Diekmann2022}. Gaussian filtering can further provide basic smoothing and noise reduction \cite{Sage2015, Ruusuvuori2010}, while more specialized filters, for instance, Laplacian of Gaussian or difference of Gaussian can improve spot contrast on non-uniform backgrounds \cite{Kong2013, Huang2015a}. Nonlinear or morphological approaches, notably median, bilateral, top-hat and $h$-dome filters \cite{Ruusuvuori2010, Huang2015a, Bright1987, Vincent217222} and wavelet domain techniques  \cite{OLIVOMARIN20021989, Hupfel2021} have also been reported to help isolate particle signals from noise. Particle localization and segmentation methods are subsequently used to identify candidate particle centers and boundaries. Localization is often achieved via template matching or cross-correlation, using a filter designed to match the expected imaging PSF (e.g., a Gaussian \cite{Stallinga2006}, corkscrew \cite{Lew:11} or a V-shaped PSF \cite{Jiang8b02800}). Segmentation relies heavily on thresholding, which can be performed globally or adaptively (locally) \cite{Li2013Fast}. Otsu's method is a widely used global approach for automatically determining an optimal threshold \cite{OtsuMethod}. More robust alternatives include minimum entropy and probabilistic thresholding \cite{Sahoo97, Hekrdla2025}. Object refinement is sometimes also needed to help separate clustered particles, a task often performed using techniques like watershedding or $h$-dome transforms \cite{Vincent217222,BARNES2014117, Rezatofighi}. 

Increasingly, ML based approaches are being adopted across the entire processing pipeline \cite{Midtvedt2021, ShuyutiParticleImpurityReview, Goyal2020, LAINE2021106077}. For particle localization and detection, ML models specifically move beyond more traditional thresholding to perform complex spatial classification and regression. Many successful architectures utilize variants of the convolutional neural network (CNN) \cite{Shiaelis2023}. Examples include the weighted average convolutional network (WAC-NET) \cite{MidtvedtACSNANO}, EfficientNET \cite{bios14080363} and specialized networks like SpotLearn \cite{Gudla2017}, DetNet \cite{Wollmann2019}, detectSPOT \cite{Mabaso2019Springer} and the deep consensus network \cite{Wollmann2021} which aggregates predictions to enhance object detection accuracy. Segmentation-focused architectures like mask R-CNN or networks employing variational splitting encoder-decoders have recently been deployed \cite{he2017maskrcnn, boyle2023iscatmaskrcnn, Ashesh2025}. U-Net architectures are also prevalent, used not only for pixel-wise semantic segmentation but also for sophisticated regression tasks. For instance, Spotiflow \cite{DominguezMantes2025} reformulates spot detection as a multiscale heatmap and stereographic flow regression problem, allowing the trained U-Net to achieve sub-pixel accurate localization. A more integrated strategy is seen in DENODET \cite{yao2024deepjointdenoisingdetection}, a specialized U-Net with a dual-decoder architecture that performs joint denoising and detection simultaneously. Further architectural developments include use of  feature pyramid based networks \cite{Song4c06395} or graph neural networks enhanced by attention-based components \cite{Pineda2023}, or morphology-guided kernel convolution \cite{Jhawar2025}. Reliance of these ML approaches on large, labeled datasets, however, frequently presents a practical bottleneck, stemming from the scarcity of high-fidelity experimental data and the challenges associated with generating accurate ground-truth annotations. Whilst this has driven efforts to develop self-supervised methods for denoising or segmentation \cite{Krull2019,Broaddus2020,Midtvedt2022}, such techniques remain prone to artifacts and hallucinations, and moreover perform poorly in low signal-to-noise regimes where distinguishing stochastic noise from faint targets is difficult without a physical prior.

A final category of methods, distinct from both filter and ML-based methodologies, are algorithms based on probabilistic models. Approaches rooted in statistical signal processing are particularly effective in our particle imaging context, as the underlying physics of image formation and the statistical properties of the noise are well characterized and can hence be explicitly modeled. Key examples include those based on maximum likelihood estimation (MLE) \cite{mortensen2010optimized} or generalized likelihood ratio tests (GLRT) for robust, threshold-free detection \cite{smith2015mboc}. Such statistical methods are also attractive due to their ability to achieve statistically optimal precision. Their performance is often benchmarked against the Cram\'er-Rao lower bound (CRLB), which defines the fundamental limit on localization precision for any unbiased estimator \cite{ober2004localization,Foreman2011}. Unlike heuristic methods, MLE-based approaches can be statistically efficient, asymptotically achieving this bound \cite{Abraham2009, Grover2010}. Furthermore, Bayesian approaches offer a framework for integrating prior knowledge to handle model uncertainty in single-molecule detection \cite{Tang2016}. Such a model-based paradigm is also particularly powerful, as it facilitates simple generalization to, for example, three-dimensional single-molecule localization or orientational studies, in which complex engineered imaging PSFs are used to encode additional particle properties \cite{Pavani2009, Foreman2008c}. Our proposed algorithm builds on these advantages and formulates particle counting as a discrete multiple-hypothesis testing problem governed by an information-theoretic complexity penalty. This offers robust, physics-interpretable counts without requiring extensive training datasets as with ML strategies, or arbitrary parameter tuning inherent to threshold and filter based methods.

\section{Methods}\label{sec:methods}

\subsection{Particle counting algorithm} \label{sec:pipeline}

\subsubsection{Analysis pipeline}
\begin{figure}[t]
  \begin{center}
  \includegraphics[width=\columnwidth]{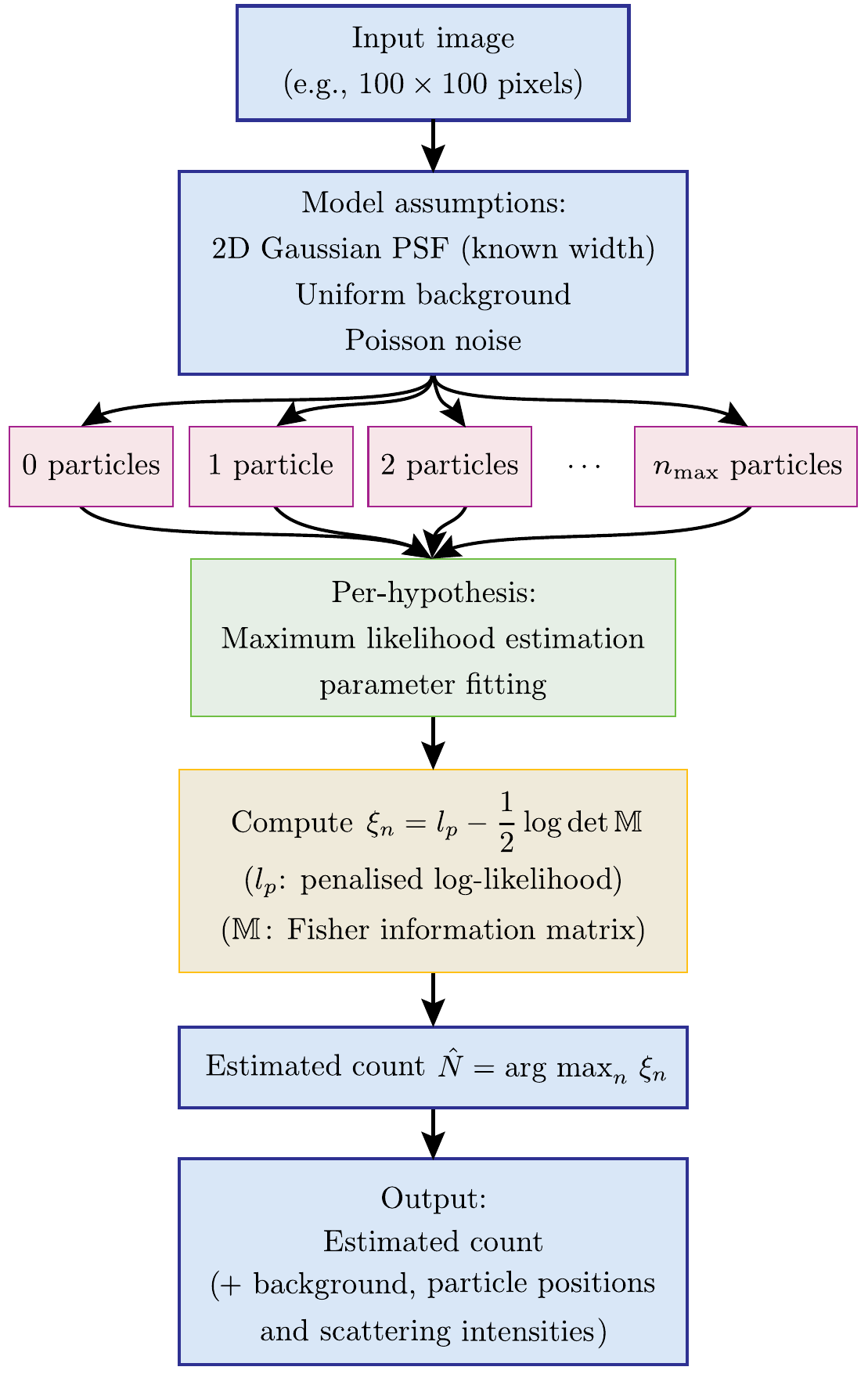}
  \caption{\textbf{Workflow of the particle detection algorithm.} 
  The input image is tested against multiple hypotheses ($H_0, H_1, \ldots, H_{n_{\text{max}}}$), corresponding to 0 to  $n_{\text{max}}$ particles, 
  each modeled as a sum of Gaussian PSFs on a uniform background. 
  For each hypothesis $H_n$, the algorithm determines the model parameters by maximum likelihood estimation, 
  computes a penalized log-likelihood ($\ell_p$) and Fisher information matrix ($\mathbb{M}$) under the assumption of Poisson noise. The corresponding scores $\xi_n = \ell_p - \tfrac{1}{2} \log\det(\mathbb{M})$ are then calculated. 
  The hypothesis with the highest $\xi_n$ is selected, yielding the estimated particle count $\hat{N}$, background level, particle positions and scattering intensities.}
  \label{fig:workflow}
  \end{center}
\end{figure}

To determine the number of particles in an image, we adopt a multiple-hypothesis testing framework. Each hypothesis corresponds to a different particle count, and the associated model is fitted to the data and evaluated using a penalized likelihood score. We use MLE to fit model parameters under each hypothesis, before selecting the hypothesis with the highest score as the final particle estimate. The bespoke analysis pipeline (written in Python and available publicly \cite{OpticalTheoryGroup_ParticleDetection}) is depicted in Figure~\ref{fig:workflow} and consists of the following steps for which further details are given in the subsequent sections:
\begin{enumerate}
    \item Coarse peak detection: Identify candidate positions using \texttt{peak\_local\_max()} from \texttt{scikit-image}.
    \item Image segmentation: Large images are divided into smaller sub-images whose size are chosen such that average particle counts are low ($\sim 0-4$).
    \item Hypothesis $H_0$: For each sub-image under a hypothesis of no particles present we estimate the background level as the mean of all pixel values (the sole model parameter), corresponding to the maximum likelihood estimate.
    \item Hypotheses $H_1, H_2, \ldots$: For each sub-image, under a hypothesis of $n$ particles present ($n\in [1,\ldots, n_{\text{max}}]$), we estimate the image background and particle parameters (scattering strength and position) by maximizing the associated (penalized) log-likelihood $l_p$, subject to a regulariser designed to constrain particle positions within the image region.
    \item All hypotheses: Compute the  Fisher information matrix $\mathbb{M}$, and the penalized score
    \begin{align}
    \xi_n = \ell_p - \tfrac{1}{2}\log \det \mathbb{M}, \label{eq:score}
    \end{align}
for which the latter term serves as an information-criterion style complexity correction \cite{schwarz1978bic,burnham2002model}.
    \item Count determination: The hypothesis with the largest score $\xi_n$ is identified and the corresponding particle count taken as the output, $\hat{N}$ of our estimation algorithm, i.e., $\hat{N} = \text{arg max}_n~\xi_n$ \cite{kay1998detection}.
\end{enumerate}

\subsubsection{Image and data model}
In this work we restrict attention to a dark-field imaging setup (see Section~\ref{sec:experimental_method}), whereby the image of a single particle will appear as a bright spot on a (ideally) low background. We model the PSF for the imaging system as a normalized Gaussian, i.e.,
\begin{equation}
f(x,y) = \frac{1}{2\pi\sigma^2}\,\exp\!\left(-\frac{x^2+y^2}{2\sigma^2}\right), 
\end{equation}
where $(x,y)$ are the image coordinates and $\sigma$ defines the PSF width. 
Although choice of a Gaussian PSF is not rigorously accurate, it enables significant computational gains whilst incurring minimal accuracy loss \cite{Stallinga2006}. The model image intensity $I(x,y)$ on the detector plane when $n$ particles are present is hence
\begin{align}
I(x,y) = I_{BG}(x,y) + \sum_{k=1}^{n} I_k\, f(x-x_k^p,\,y-y_k^p),
\end{align}
where $I_{BG}$ is the background intensity profile, and $I_k, (x_k^p, y_k^p)$ are the amplitude and position of the $k$-th particle. Note that the form of $f(x,y)$ ensures that $\int f(x,y)\,dx\,dy = 1$, such that the particle amplitude $I_k$ directly represents the total expected intensity contribution from particle $k$.

Modern imaging setups employ pixelated detectors, i.e. a camera. The total signal intensity captured by a single pixel $j$ on the camera is given by
\begin{align}
\lambda_j = \lambda_{BG} + \sum_{k=1}^m I_k f_j(x_k^p,y_k^p), \label{eq:model_int}
\end{align}
where we have assumed the background signal $\lambda_{BG}$ is independent of pixel across each of the small sub-images (although may vary between sub-images), $f_j(x_k^p,y_k^p) = \iint_{\Omega_j} f(x-x_k^p,y-y_k^p) dx dy$  and $\Omega_j$ denotes the spatial domain of the $j$-th pixel. Henceforth we will express positions in terms of pixel widths, such that $\Omega_j$ defines the domain $\{x,y\, | \, |x-x_j| < 1/2,  |y-y_j| <1/2 \}$. Each sub-image will be assumed to have normalised coordinates such that $\{x,y \,| -1/2 \leq x \leq w_x-1/2, -1/2\leq y \leq w_y -1/2 \}$, where $w_x$ and $w_y$ are the dimensions of the sub-image (in pixels). The fraction of the total scattered signal from a particle falling onto a single detector pixel depends on the relative alignment of the PSF to the pixel array. For later convenience we define the maximum fraction (corresponding to when the PSF is centered on a pixel - see Supplementary Material for further detail) to be $f_{\text{max}}$.

\subsubsection{Penalized log-likelihood function}
 We assume that noise present on the model image intensity follows a Poisson distribution (with mean equal to the model value, i.e. Eq.~\eqref{eq:model_int}), since for small weakly scattering nanoparticles, the photon count per pixel is low. Operation in such a quantum noise limited regime is routinely achieved in single particle and fluorescence imaging \cite{Mabaso_Withey_Twala_2018, Smal2010}. Assuming  that noise in each pixel is independent, the likelihood function is therefore
\begin{equation}
    \mathcal{L}(\mathbf{v} \mid \boldsymbol{\lambda}) = \prod_{j=1}^{M} \frac{\lambda_j^{v_j} e^{-\lambda_j}}{v_j!},
\end{equation}
where $M$ is the total number of pixels in a given sub-image, $\mathbf{v} = [v_1,v_2,\ldots, v_M]$ is the (vectorized) image data composed of  the observed intensities $v_j$ at each pixel $j$, and $\boldsymbol{\lambda} = [\lambda_1, \lambda_2,\ldots, \lambda_M]$ is the expected image intensity. 

Neglecting the model independent $\sum_j\log(v_j!)$ term we can form the log-likelihood function
\begin{equation}
\ell(\boldsymbol{\theta}) = \sum_{j=1}^{M} \bigl[ v_j \log \lambda_j(\boldsymbol{\theta}) - \lambda_j(\boldsymbol{\theta}) \bigr],
\end{equation}
where $\boldsymbol{\theta} = [\lambda_{BG},I_1,x_1,y_1,\ldots,I_n,x_n,y_n]$ denotes the vector of model parameters to be found (i.e., background intensity, particle positions and scattering strengths).  

To prevent particle position estimates from drifting outside the image boundary, we add a regularizer $\mathcal{P}(\boldsymbol{\theta})$ to the log-likelihood. 
The resulting penalized objective is
\begin{equation}
\ell_p(\boldsymbol{\theta}) = \ell(\boldsymbol{\theta}) - \alpha \mathcal{P}(\boldsymbol{\theta}), \label{eq:lp}
\end{equation}
where $\mathcal{P}(\boldsymbol{\theta})= \sum_{k=1}^N [\mathcal{P}_x({x_k^p}) + \mathcal{P}_y({y_k^p})]$, and
\begin{equation}
\mathcal{P}_s(t) =
\begin{cases}
(t+0.5)^q  & t < -0.5, \\
(t - w_s + 0.5)^q & t \geq w_s - 0.5 ,\\
0 & \text{otherwise}.
\end{cases}
\end{equation}
 The penalty $\mathcal{P}(\boldsymbol{\theta})$ is thus zero when all particle positions lie within the valid region, and is  activated only if a fitted position crosses the image edge. Note that, $q$  and $\alpha$ are free parameters that can be selected by the user. Our choice of a polynomial penalty yields smooth first and second derivatives. The solver (see below) could hence be provided with closed-form expressions for the Jacobian vector ($\mathbf{J}$) and Hessian matrix ($\mathbb{H}$) of $-\ell_p(\boldsymbol{\theta})$ with respect to $\boldsymbol{\theta}$ to help ensure stable convergence. Full expanded expressions for $\mathbf{J}$ and $\mathbb{H}$ are included in the Supplementary Information. 

\subsubsection{Maximum likelihood estimation}
To find the maximum likelihood estimate of the model parameters $\boldsymbol{\theta}$ we numerically maximize the penalized log-likelihood function (Eq.~\eqref{eq:lp}) using a Newton-type optimizer.  Specifically we supplied $-\ell_p(\theta)$ to SciPy’s \texttt{trust-exact} minimizer so as to maximize $\ell_p(\theta)$.  For our study we choose $\alpha = 10^5$ to match the scale of our modeled signal intensities and provide rapid convergence. A cubic penalty, $q=3$, was chosen because its gradient  increases quadratically with distance outside the image boundary, such that larger excursions are corrected more aggressively than a more conventional quadratic penalty. This helps to prevent the minimizer from lingering outside the valid region for too many iterations if a large step is taken.

\subsubsection{Hypothesis testing}
We consider a range of hypotheses, $H_n$, for $n=0,1,\ldots n_{\text{max}}$, where $H_n$ assumes an image contains $n$ particles. For each hypothesis, MLE of the model parameters $\boldsymbol{\theta}$ is performed as described above and the penalized score, given by Eq.~\eqref{eq:score} calculated. For all tested hypotheses, we assign a score according to 
\begin{align}
\xi_n = l_p - \frac{1}{2}\log\det{(\mathbb{M}}).
\end{align}
The $\frac{1}{2}\log\det{(\mathbb{M}})$ penalty term represents an information-criterion style complexity correction applied to guard against overfitting in higher order hypotheses which possess more degrees of freedom \cite{schwarz1978bic,burnham2002model}.  This penalty term is based on the minimum description length principle, which is asymptotically equivalent to the referenced Bayesian information criterion (BIC) and provides a rigorous foundation for balancing model accuracy with complexity \cite{kay1998detection}. The hypothesis found to possess  the greatest score $\xi_n$ is selected as the resulting estimate $\hat{N} = \text{arg max}_n~\xi_n$ of the true underlying particle count $N$. The maximum number of particles to consider, $n_{\text{max}}$, can be set dynamically, e.g., based on identification of turning points in $\xi_n$ as a function of $n$. Alternatively, as in the case of our study, $n_{\text{max}}$ can be set statically based on \emph{a priori} estimates of mean particle densities. 
In particular, for an (estimated) average particle density $\meanN$ we can select $n_{\text{max}}$ such that 
\begin{align}
p_{N}(N > n_{\text{max}}) = 1- \sum_{k=0}^{n_{\text{max}}} \frac{\bar{N}^k}{k!} \exp[-\bar{N}]  < \varepsilon, 
\end{align}
i.e., the probability that the actual count of particles is larger than $n_{\text{max}}$ (assuming a Poisson point process) lies below some threshold $\varepsilon$. Note that when the true count $N$ is greater than $n_{\text{max}}$ we empirically observed that the corresponding particle count was estimated as $n_{\text{max}}$. Theoretically, this would be expected because the likelihood increases with increasing $n$ for $n<N$.

\subsubsection{Initialization strategy}
Suitable parameter initialization (i.e. of $\boldsymbol{\theta}$) is important for stable convergence in our MLE. To this end, we set the initial parameter values as follows:
\begin{itemize}
    \item Background: $\lambda_{BG}$, the uniform background intensity, is initialized as the minimum pixel value in the image.
    \item Particle intensities: $I_k = \Delta I \cdot 2\pi\sigma^2$ for all candidate particles, 
          where $\Delta I = \max(\mathbf{v}) - \min(\mathbf{v})$.
    \item Particle positions: $(x_k^p,y_k^p)$ are taken from peaks identified by \texttt{peak\_local\_max()}, 
          ordered by descending peak intensity, up to the assumed number of particles $n$ in the current hypothesis.
\end{itemize}

\subsection{Synthetic image sets for performance evaluation}

Synthetic image data provide a uniquely controllable testbed for evaluating the performance of our algorithm since ground-truth parameters, including the particle numbers, are known exactly. By benchmarking algorithm results against fully defined ground-truth datasets, we can directly quantify the algorithm’s accuracy under systematically varied conditions. In this vein we performed two types of parametric performance studies to analyze the counting accuracy and resolvability of particles respectively. 

\subsubsection{Counting accuracy test set} \label{sec:synthetic}
We generated a set of $100\times100$ pixel (sub-)images containing 0 to 4 particles to test our detection algorithm. For each particle count, we created a stack of 10,000 images using a predefined PSF width, background level, and particle intensity. Particles were randomly distributed within the image. To isolate the effects of imaging parameters on algorithm performance relative to boundary effects, particle positions were restricted to regions in which the full PSF fell within the image. Pixel values were generated by Poisson sampling from the expected model intensity, thereby capturing fundamental photon shot noise inherent in optical detection \cite{ober2004localization, hasinoff2014photon,thompson2002precise}.

\begin{table}[b]
\begin{center}
\caption{Baseline parameter values used for generation of simulated image datasets.\label{tab:baseline}}
\begin{tabular}{|l|c|c|}
\hline
\textbf{Parameter}& \textbf{Symbol} & \textbf{Baseline}\\
\hline\hline
Image dimensions & $w_x=w_y$ &$100$ pixels\\\hline
PSF width & $\sigma$ & $2$ pixels\\\hline
Background count & $\lambda_{BG}$ & $2000$\\\hline
Particle scattering intensity & $I_k$ & $20,000$\\\hline
Signal-to-background ratio & SBR & $\approx 0.39$ \\\hline  
Signal-to-noise ratio & SNR& $\approx 14.8$  \\
\hline \end{tabular}
\end{center}
\end{table}

Baseline simulated imaging conditions are summarised in Table~\ref{tab:baseline}. Relative to these baselines values we generated further image sets for variations in the particle scattering intensity $I_k$ (assumed the same for all particles) and pixel background $\lambda_{BG}$, which were sampled logarithmically in the ranges $1/64\times$ to $8\times$ and $1/2\times$ to $2048\times$ respectively. Noting that the signal-to-background  ratio (SBR) is here defined as
\begin{align}
    \text{SBR} = \frac{I_k f_{\text{max}}}{{\lambda_{BG}}},
\end{align}
the variation in  $I_k$ directly corresponds to variation of the SBR from  $1/64\times$ to $8\times$ the baseline value of $\approx 0.39$. Meanwhile, changes in $\lambda_{BG}$ correspond to variation of the SBR from $2\times$ to $1/2048\times$ baseline.

To quantify the effect of the signal-to-noise ratio, defined here as
\begin{align}
    \text{SNR} = \frac{I_k f_{\text{max}}}{\sqrt{I_k f_{\text{max}}+\lambda_{BG}}},
\end{align}
to reflect the Poisson distributed detection noise, we generated further synthetic images in which both the particle intensity and the background level are multiplied by a common scale factor $s$. This preserves the SBR ratio while increasing the absolute photon counts per pixel and the SNR by a factor of $\sqrt{s}$. Scaling factors (again logarithmically spaced) ranging from $1/16$ to $8$, corresponding to relative SNRs of $1/4$ to $2\sqrt{2}$ were chosen.

Finally, we also generated two further test image sets corresponding to different optical magnifications. As imaging magnification increases, we assume PSF spreads across more pixels while conserving the total photon signal. Accordingly, the peak pixel intensity decreases while the integrated signal per particle is unchanged. The two distinct datasets compared two background models: type~I, an optical background that dilutes with magnification in the same way as the signal ($\propto 1/\text{magnification}^2$), and type~II, a static non-optical background that is  independent of magnification. Physically, the type I model is more applicable when the background signal arises from the sample or illumination optics (e.g. background fluorescence or stray light), whereas the magnification independent type II model describes signals originating at the detector, such as thermal counts and read-out noise. 

\subsubsection{Particle resolvability test set}\label{sec:resolvabilitydata}
Optical imaging systems are intrinsically limited in resolution due to the diffraction of light. Consequently the image of two closely spaced point-like objects differs only marginally from that of a single particle. Although in a noise-free system two point objects can theoretically be resolved down to infinitely small separations, e.g., by deconvolving the image with the known imaging PSF, in reality, two closely spaced particles can be difficult to discriminate since noise limits the minimum detectable intensity difference. To study  the ability of our algorithm to resolve closely spaced particles in the presence of noise, we generated an additional set of $100\times100$ pixel images containing exactly two particles, with inter-particle distance, $d$, ranging from 0 to $6\sigma$ in increments of $0.2\sigma$. The assumed PSF width $\sigma$ was additionally varied from 1 pixel to $4\sqrt{2}$ pixels. 
For each distance and PSF width combination, 10,000 images are produced. The inter-particle orientation was randomized for each image, and the centroid position was placed uniformly at random within $\pm 0.5$ pixels of the image center along both axes to remove any pixel-array alignment effects, such as informational oscillations or biases \cite{Torok2019}. Note, that we also held the SBR constant for different PSF widths through appropriate scaling of particle scattering strength.

\subsection{Experimental image set}\label{sec:experimental_method}

\subsubsection{Dark field imaging setup}

To evaluate the algorithm under real imaging conditions, we applied it to a dark-field microscopy dataset of metallic nanoparticles. We illuminated the nanoparticles using broad-spectrum LEDs (Cree C503D-WAN, 40k mcd intensity) at a large oblique angle, such that only the scattered light was collected by a low-magnification microscope objective (Olympus LUCPlanFL N 20x, 0.45 NA) and the image subsequently projected onto a color RGB camera (Teledyne FLIR Grasshopper 3, GS3-U3-89S6C-C, Sony IMX255 sensor). Piezo stages (Thorlabs PD2/M 5 mm Linear Stage with Piezoelectric Inertia Drive) allowed us to automate the system and capture multiple images at distinct sample regions. 

\subsubsection{Coronavirus assay}
We attempted to approximate genomic RNA detection by mixing two single-stranded DNA (ssDNA) sequences; a 188-bp ssDNA (``E-DNA") derived from SARS-CoV-2 Envelope gene (\textit{E} gene) or a 200-bp ssDNA (``N-DNA") from SARS-CoV-2 Nucleocapsid gene (\textit{N} gene). When the ssDNAs bind to nanoparticles with complementary DNA handles, it leads to the formation of nanoparticle clusters.
Following \cite{Jhawar2025} we used a mixture of nanoparticles in this experiment; gold-shell silica-core nanoparticles (80~nm diameter silica + 20~nm thick Au shell), 25~nm radius spherical gold nanoparticles and 25~nm radius spherical silver nanoparticles respectively. Details about the surface functionalization can be found in the Supplementary Information. The nanoparticles were mixed with a buffer formulated to reduce non-specific aggregation, after which the target DNA (or blank buffer) was added at a concentration of 10 pM. The samples were briefly heated for 15 min at 50~$^\circ$C before transfer to a home-made imaging chamber for dark-field imaging. We considered three distinct concentrations, specifically
\begin{enumerate}
    \item \textbf{Low}: Au nanoshell: 0.3~pM, Au and Ag spheres: 1.5~pM each. Total:  3.3~pM.
    \item \textbf{Medium}: Au nanoshell: 0.6~pM, Au and Ag spheres: 3~pM each. Total: 6.6~pM.
    \item \textbf{High}: Au nanoshell: 1.2~pM, Au and Ag spheres: 6~pM each. Total: 13.2~pM.
\end{enumerate}
Note that since the nanoparticle samples used comprised of a mixture of different species possessing distinct scattering cross-sections, we consequently introduced inter-particle variability in $I_k$ thereby providing an additional test of the robustness of our algorithm. In principle, the distinct spectral response of each nanoparticle species can provide additional information about the content of observed clusters, however, we did not exploit such spectral information in this study.


\section{Results and discussion}


\begin{figure}
  \begin{center}
  \includegraphics[width=\columnwidth]{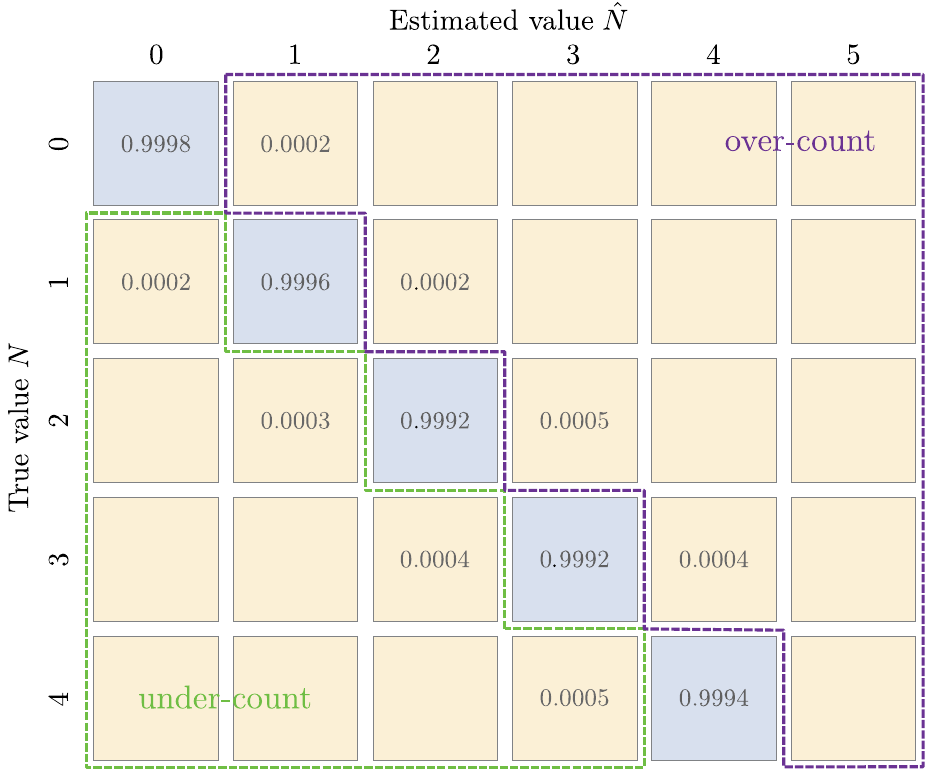}
\caption{\textbf{Baseline confusion matrix.} Confusion matrix for the particle count estimation algorithm under baseline simulation conditions (Table~\ref{tab:baseline}), illustrating the probability of the estimated count $\hat{N}$ (columns) given the true count $N$ (rows). Strong diagonal dominance is evident reflecting minimal over- or under-counting.}
\label{fig:confusion}
  \end{center}
\end{figure}
\begin{figure*}
  \begin{center}
  \includegraphics[width=\textwidth]{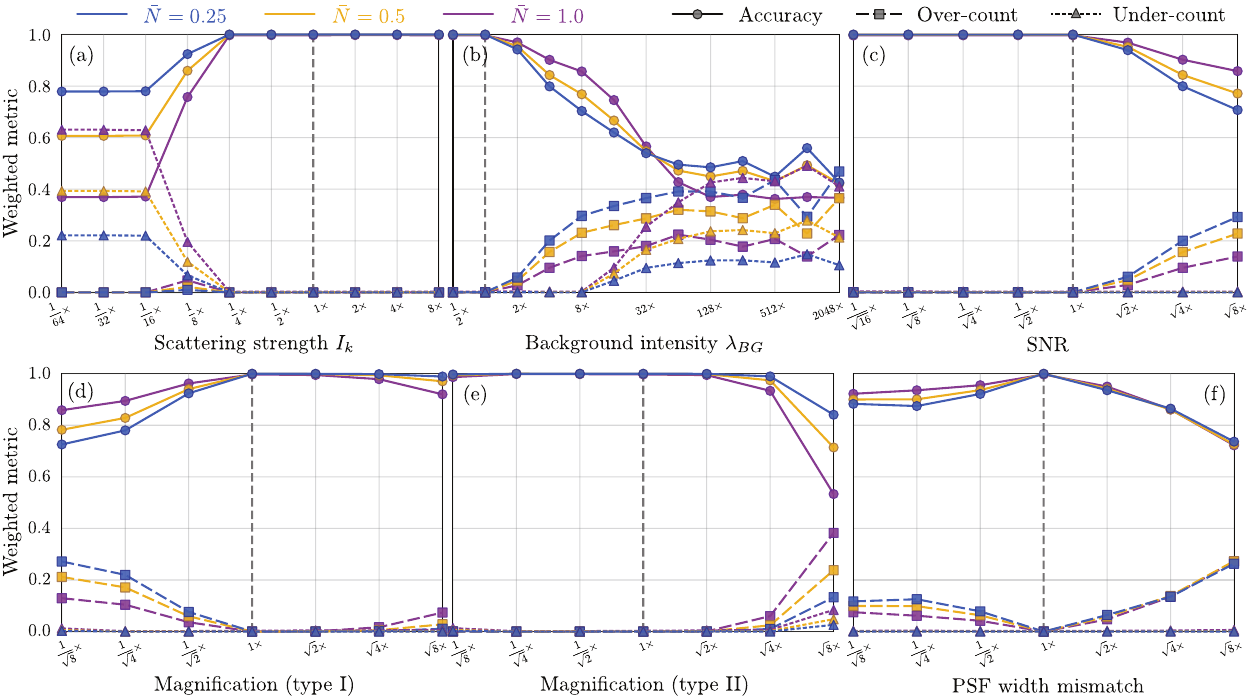}
\caption{\textbf{Weighted accuracy under simulated conditions.} Weighted accuracy (solid lines), over-count rate (dashed) and under-count rate (dotted) versus 
(a) particle scatterong strength, (b) background intensity, (c) relative SNR at fixed SBR, (d) magnification (type~I), where both signal and background scale as $1/\text{magnification}^2$ (optical stray-light–dominated),
(e) magnification (type~II), where background remains constant (sensor/electronic-noise–dominated),
and (f) the ratio of the assumed PSF width used in MLE to the true PSF width. 
Each panel shows results for  average particle densities per image ($\meanN$) of 0.25 (blue), 0.5 (orange), and 1.0 (purple). Each point aggregates outcomes from 10{,}000 test images per true count (0--4). 
Baseline conditions are given in Table~\ref{tab:baseline} and depicted by vertical dashed gray lines. Hypotheses up to five particles were evaluated in all cases.}
\label{fig:accuracy_results}
  \end{center}
\end{figure*}

\subsection{Counting accuracy}\label{sec:accuracy_results}
This section describes the performance of our proposed hypothesis based particle counting algorithm based on six parametric studies using synthetic datasets. Specifically, we varied particle scattering strength, image background intensity, relative SNR and optical magnification (type I and type II), as described in Section~\ref{sec:synthetic}. Additionally, using the baseline imaging dataset, we varied the width of the PSF used in model fitting relative to the true underlying PSF. For each ground-truth particle count $N$, we recorded the relative frequency $\nu_{N}(\hat{N})$ of each estimated particle count, $\hat{N}$, output as a confusion table (the baseline confusion table is presented in Figure~\ref{fig:confusion}, with further examples provided in the Supplementary Information) where $\sum_{\hat{N}} \nu_N(\hat{N}) = 1$. Although our datasets contained images with a maximum of four particles, hypotheses up to $H_5$ were tested to help us assess any tendency of the algorithm to over-count. Such misclassifications were negligible, confirming that the algorithm does not spuriously inflate counts under the tested densities.

To reflect the distribution of expected counts inherent in a Poisson counting process, we evaluate overall algorithm performance using a weighted accuracy defined as
\begin{align}
\mathcal{A} = \sum_{N=0}^{4} p_N(N;\meanN)\, \nu_{N}(N), \label{eq:weighted_accuracy}
\end{align}
where $p_N(N;\meanN)$ is the Poisson probability mass function for $N$ particles with mean $\meanN$, $\nu_{N}(N)$ is the fraction of correct predictions and the summation is performed over the leading diagonal of the confusion matrix (blue boxes in Figure~\ref{fig:confusion}). An unweighted accuracy, in contrast, treats rarer high-count cases as equally important as common low-count ones, ignoring the fact that counts near the mean occur far more often in practice. Adopting a weighted approach, moreover, allows us to realistically assess algorithm performance for differing particle densities by adjusting the relative weights in the averaging process. Specifically, we report results for three  choices of $\meanN$ ($0.25, 0.5,$ and $1.0$ particles per image). In so doing, we can gain extra insight into algorithm performance. If, for example, errors are concentrated in lower-count cases (e.g., $N=0$ or $1$), then the lower-density test (e.g., $\meanN=0.25$) will show lower weighted accuracy than a higher-density test (e.g., $\meanN=1.0$), since low counts dominate the Poisson mix at small $\meanN$. Conversely, if errors arise mainly at higher counts (say $N \ge 2$), weighted accuracy will decline as $\meanN$ increases. In this way, the trend of accuracy versus $\meanN$ indicates which count regimes contribute most to errors. We also similarly define weighted over-count and and under-count rates corresponding to Poisson weighted sums over the purple and green regions depicted in Figure~\ref{fig:confusion}. Summary plots of the resultant weighted performance metrics are given in Figure~\ref{fig:accuracy_results}. 

\subsubsection{Effect of particle scattering strength}
Figure~\ref{fig:accuracy_results}(a) shows the observed weighted accuracy (solid lines) as a function of scatterer strength $I_k$ with the background held fixed at 2{,}000 and $\sigma=2$~pixels, along with the weighted over-count (dashed) and under-count (dotted) failure rates.
At the shared baseline condition ($I_k=20{,}000$ for all particles $k$), weighted accuracy is essentially 1.0, serving as the high-quality reference. Reducing scatterer strength decreases the per-particle intensity while the background remains unchanged, which lowers the peak-pixel fraction relative to background.
For $\sigma=2$~pixels the central pixel contains at most $f_{\max}\approx 3.90\%$ of the particle intensity (see Supporting Information). This corresponds to at most 780 counts at $1\times$, 390 counts at $1/2\times$, and 195 counts at $1/4\times$, all against a background mean of 2{,}000.
Despite the central pixel being less than 10\% of the background in the $1/4\times$ case, accuracy remains essentially unity down to this regime and only begins to deteriorate for weaker scattering, where a rise in under-count rate is evident. The plateaus on the far left of Figure~\ref{fig:accuracy_results}(a) correspond to regimes where the algorithm predicts zero particles for nearly all images, that is to say when Eq.~\eqref{eq:weighted_accuracy} reduces to $\mathcal{A} = p_N(0;\meanN) = \exp(-\meanN)$, corresponding to $\approx 0.368$, 0.607 and 0.779 for means of 1, 0.5 and 0.25 respectively.

\subsubsection{Effect of background level}
In Figure~\ref{fig:accuracy_results}(b) it is observed that weighted accuracy decreases monotonically with increasing background signal $\lambda_{BG}$. At the baseline background of 2{,}000 counts, accuracy is again essentially unity, however, as the background rises, the peak-to-background ratio shrinks and accuracy steadily declines. When the background reaches approximately $128\times$ the baseline, the usable contrast is so low that the estimator effectively toggles between the background-only ($\hat{N}=0$) and single-particle ($\hat{N}=1$) models; predictions with $\hat{N}\geq 2$ do not occur in this regime (see the Supplementary Information for supporting confusion matrices and further discussion).

Reducing scattering strength and increasing background both lower the SBR. In our simulations, the weighted accuracy in Figure~\ref{fig:accuracy_results}(a) remains near the baseline down to $\sim 1/4\times$ particle scattering strength $I_k$ (SBR reduction of $\sim 4$), whereas in Figure~\ref{fig:accuracy_results}(b) a visible decline appears at  $2\times$ baseline background (SBR reduction of 2). Given the two parametric studies probe distinct parameter regions, similar SBR `thresholds' would not be expected. 

\subsubsection{Effect of SNR}
When scaling both particle intensity and background by a common scale factor $s$ (see Section~\ref{sec:synthetic}), thereby fixing SBR whilst scaling SNR by $\sqrt{s}$, we observe that the weighted
accuracy is essentially unity across a broad range of scale factors (see Figure~\ref{fig:accuracy_results}(c)). 
At very large $s$, however, accuracy shows a modest but reproducible decline. 
This behavior is counterintuitive since one might expect that ever-higher SNR should only improve detection. 
Instead, the elevated photon counts amplify the absolute magnitude of Poisson fluctuations, which can generate spurious maxima resembling weak narrow PSFs. The dominant error mode is misclassification of true $N=0$ images as possessing a single particle ($\hat{N}=1$) as reflected in the increasing over-count rate. Thus, while performance remains robust across most of the tested range, extremely high SNR introduces fluctuation-driven artifacts that offset the expected gains. This  response reflects algorithm-specific sensitivity to the structure of the likelihood surface.

\subsubsection{Effect of system magnification}
Under type~I conditions varying magnification of an imaging system affects both the PSF width (thereby affecting per-pixel counts of scattered photons) and the background level. We find  that in this case weighted accuracy remains constant across a range of PSF widths from $\sim 1/\sqrt{2}$ to $\sim {2}\times$  the baseline, but then reduces at both lower and higher magnifications (see Figure~\ref{fig:accuracy_results}(d)). This trend is consistent with results found in single molecule localization studies, which demonstrated that in ideal conditions localization precision can be magnification independent, but pixelation and noise impose a practical optimum \cite{ober2004localization}. At low magnification the (narrow) PSF is under-sampled, which in our tests primarily produced rare misclassification of true $N=0$ images as $\hat{N}=1$, i.e. false positives exhibiting as over-counting (see Supplementary Information for corresponding confusion matrices). On the other hand, at high magnification, the PSF spans many pixels. The increased number of pixels, combined with larger noise fluctuations can produce spurious maxima which are subsequently identified as additional particles (over-counting). Moreover, when multiple particles are present, broader PSFs can result in significant spatial overlap such that close particle pairs are not clearly resolved (again due to noise) and hence are only counted as a single particle (weak under-counting). Note that the latter primarily affects larger $N$ cases and therefore does not significantly affect the weighted under-count rate, but is more evident in the corresponding confusions tables (see Supplementary Information).

Under type~II conditions (variable PSF width, but fixed background intensities), weighted accuracy stays high at magnifications up to $\sim 2\times$ baseline, but drops once magnification is large enough that the fixed background becomes significant relative to the diluted per-pixel signal (Figure~\ref{fig:accuracy_results}(e)). This is analogous to the readout-noise-dominated regime discussed in \cite{ober2004localization}, where increasing magnification spreads the signal without reducing the fixed noise floor. No decline in accuracy is seen at smaller PSF widths (lower magnifications). 

Realistically, both type~I and type~II type backgrounds are likely to be present in an imaging system, such that a practical magnification window is expected. The lower end of this window is determined by type~I effects, whilst the  the upper end is dictated by both type I and II effects. The dominant factor at high magnifications will depend on whether SBR or SNR limit performance whereby either the fixed noise floor penalizes very high magnification or large noise fluctuations produce spurious maxima and degrade particle resolvability respectively. Empirically, we therefore note that best performance is obtained when the PSF spans approximately 2-4 pixels, which is consistent with behaviour reported in the literature \cite{Loretan2025}.

\subsubsection{Effect of PSF mismatch}
The analysis presented thus far assumed that the PSF width used in analysis matched the true PSF width $\sigma$. Precise determination of the true PSF width may be practically difficult and so we also evaluate the performance of the counting algorithm under baseline conditions albeit with a mismatch in the model  PSF used in our MLE relative to the true $\sigma$. The  PSF mismatch was varied from $1/\sqrt{8}\times$ to $2\sqrt{2}\times$ the baseline truth. Figure~\ref{fig:accuracy_results}(f) shows that such PSF mismatch affects weighted accuracy across the entire simulated parameter range. Performance tended to plateau when the analysis PSF was narrower than the true one, while it dropped off more sharply when the analysis PSF was broader, a trend that would likely continue beyond the tested range. In both cases the principal failure mode was over-counting. Nevertheless, declines in accuracy were limited and notably approximately independently to $\meanN$, hence demonstrating that the method is relatively insensitive to such PSF discrepancies.

\begin{figure}[t]
  \begin{center}
  \includegraphics[width=0.99\columnwidth]{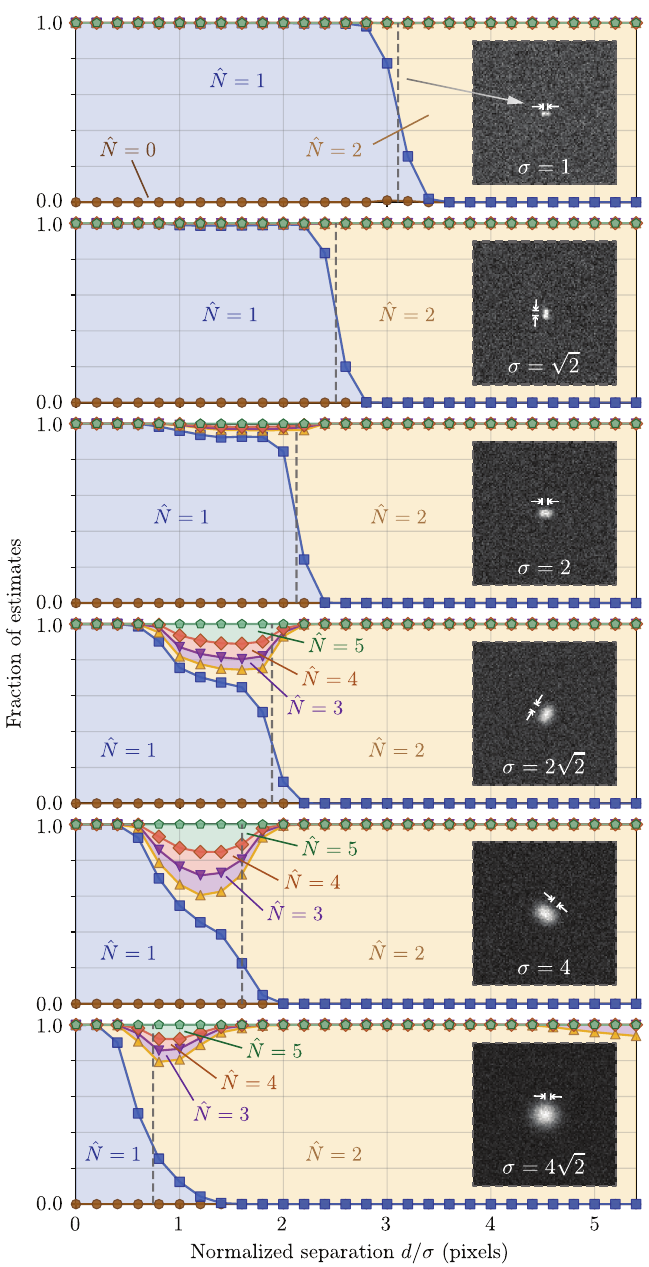}

\caption{\textbf{Two-particle resolution}. Stacked plot for the fraction of estimation outcomes $\hat{N}$ for a true two-particle image ($N=2$), as a function of normalized separation for PSF widths of $\sigma = 1$ (top panel), $\sqrt{2}, 2, 2\sqrt{2}, 4$ and $4\sqrt{2}$ pixels (bottom), each aggregated over 10,000 simulations. At large separations, correct two-particle estimates dominate. At small separations, under-counting is the primary error mode. Over-counting arises only near the transition region. Example particle images at the resolution limit (vertical dashed line corresponding to the fraction of $\hat{N}=2$ estimates falling to 50\%), for each PSF width are also shown. White arrows on inset denote corresponding threshold separation.
\label{fig:resolvability}} 

  \end{center}

\end{figure}

\subsection{Particle resolvability}\label{sec:resolvability_results}
As discussed in Section~\ref{sec:resolvabilitydata} a common problem encountered in image based particle counting is the need to resolve two particles that lie in close proximity to each other. This behaviour is similarly, unavoidable in our approach and is manifest as potential under-counting of particle numbers (which we shall term coincidence loss). Figure~\ref{fig:resolvability} shows the results of analysis of the dataset described in Section~\ref{sec:resolvabilitydata} using our hypothesis-based algorithm. Specifically, the relative frequency of each selected hypothesis (i.e., the determined particle count) is presented as a stacked plot as a function of the ground-truth particle separation $d$ (normalized to PSF width $\sigma$). Initially considering Figure~\ref{fig:resolvability}(a), we observe a sharp crossover behavior whereby at larger separations ($\gtrsim 3\sigma$ or equivalently $\gtrsim 3$ pixels in this case) two particles are correctly discriminated, however at small separations the estimator predominantly under-counts (i.e., $H_1$ wins over $H_2$ in the scoring, meaning two particles are reported as one). We define the transition point, or resolution threshold, as the separation corresponding to where the \emph{two} particle count drops to $50\%$ of the total as is indicated by the vertical gray dashed line in Figure~\ref{fig:resolvability}(a). As the PSF width ($\sigma$) is increased (panels (b)--(f)), the normalized threshold separation is seen to shift to smaller multiples of $\sigma$, whilst the absolute threshold separation naturally increases as would be expected (see Supplementary Information). We attribute the ability of our  algorithm to better resolve relatively closer pairs when the PSF is broader, to the fact that for wider PSFs information about each particle is encoded in more pixels, each of which, in the Poisson noise regime, individually exhibits a better SNR. Accordingly smaller image differences can be discerned enabling better resolvability. It should however be noted that for broader PSFs, the transition behavior is more complicated than a crossover between the estimator choosing $H_1$ or $H_2$. In particular, since PSFs are spread over multiple pixels, noise can introduce spurious maxima, as seen with the counting accuracy tests,  which are misclassified as multiple particles.  Indeed, spurious maxima become more problematic in the resolution tests than in the earlier accuracy tests, since the presence of two overlapping PSFs generates larger scattered intensities than the single particle case, such that intensity variance on each pixel is similarly larger. Such over-counting can be seen in Figure~\ref{fig:resolvability}(d)--(f) in which small contributions from 3, 4 and 5 particle estimates are seen. Overall, the algorithm however predominantly fails conservatively, underestimating rather than inflating counts, with over-counting restricted to the transition region. These error patterns are consistent with the general challenges documented in localization reviews \cite{small2014accurate}, where overlapping emitters and high background levels are recognized as primary obstacles.

Finally, we note that at the densities used in our image datasets ($\meanN = 0.25, 0.5$ or $1$), the probability of any particle overlap in a random image is only $\sim$0.3\%. Even when restricting consideration to images with at least two particles, the probability of overlap remains $\sim$1.1\% (See Supplementary Information for a derivation). These values indicate that coincidence losses are rare in the accuracy tests discussed above.

\subsection{Digital coronavirus assay} \label{sec:exp_results}
\begin{figure*}[t]
  \begin{center}
  \includegraphics[width=\textwidth]{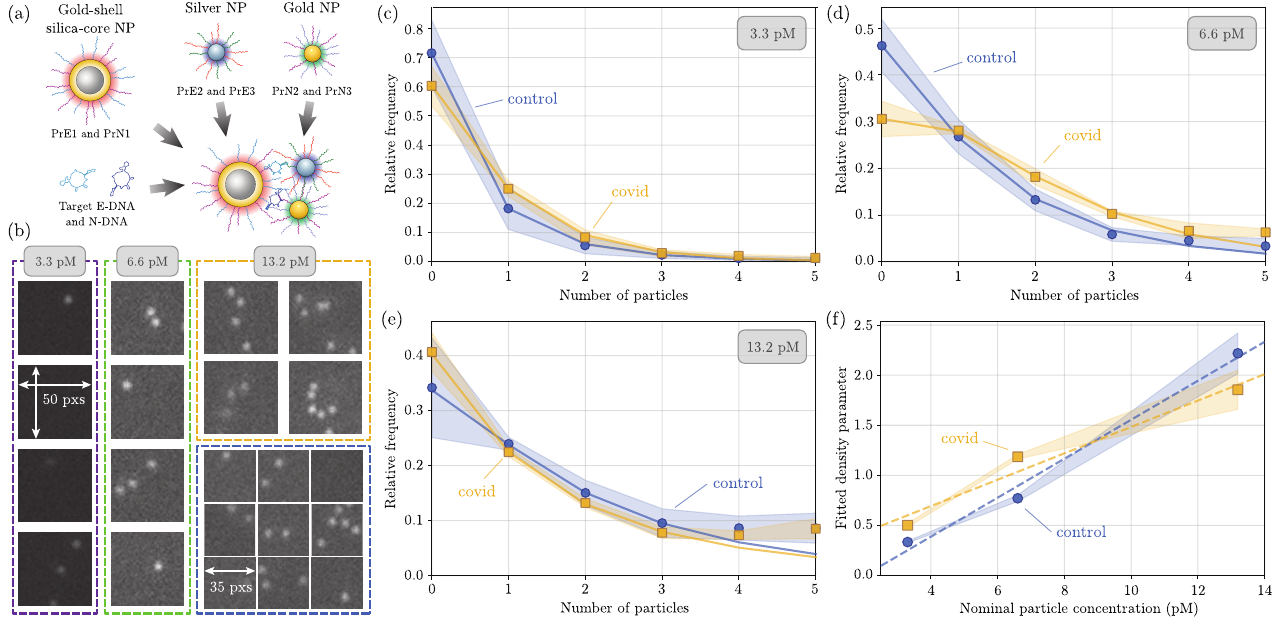}
  \caption{\textbf{Nanoparticle imaging based SARS-CoV-2 assay.} (a) Schematic of nanoparticle functionalisation (see Supplementary Information) and target induced nanoparticle clustering. (b) Example dark-field images of deposited nanoparticles for different nanoparticle concentrations. For each concentration, images shown correspond to a single $50\times 50$ pixel sub-region input into the particle counting algorithm. High nanoparticle concentrations (13.2~pM), however can frequently produce high numbers of particles ($>n_{\text{max}}$) in a given sub-image (orange dashed bounding box), such that smaller $35\times35$ pixel regions were used (blue dashed bounding box).
   (c)-(e) Distribution of estimated particle counts across all sub-images, for both control (blue markers) and SARS-CoV-2 (`covid') derived biomarkers positive (orange markers) samples, overlaid with GPD fits (solid curves). Error bands correspond to inter-image standard deviations. (f) Fitted density parameters (per $50\times50$ pixel region) for different nanoparticle concentrations as found from GPD fitting (note, the 13.2~pM density was rescaled) shown with corresponding linear fits (dashed curves).}
  \label{fig:exp}
  \end{center}
\end{figure*}

Finally, in this section we consider real world application of the proposed particle counting algorithm to experimental nanoparticle images taken as part of a bio-assay for detection of SARS-CoV-2 DNA biomarkers (see Section~\ref{sec:experimental_method}). Specifically, presence of the target DNA markers induces nanoparticle clustering as depicted in Figure~\ref{fig:exp}(a) \cite{Jhawar2025}. Given the short incubation times used, nanoparticle aggregate sizes remained below the optical diffraction limit, such that, like individual nanoparticles, clusters appear as single bright spots in the experimental images. Analyte presence can nevertheless be detected through quantification of the apparent nanoparticle concentration (or more strictly of the bright spots), $\meanN$, observed in the acquired dark field images. Acquired images (2448 × 2048 pixels) were converted to gray-scale and initially cropped by a factor of 0.7 along each dimension to reduce illumination and imaging heterogenity across the field of view. Images were then subsequently divided into non-overlapping pixel sub-images, each of which were processed independently using the proposed algorithm to estimate the corresponding particle count. For the low (3.3~pM) and medium (6.6~pM) nanoparticle concentration experiments, $50\times 50$ sub-image sizes were selected so as to ensure images containing $n_{\text{max}}=5$ or more particles were rare (see Figure~\ref{fig:exp}(b) for examples of typical sub-images). For the higher concentration sample (13.2~pM) use of $50\times50$ images sizes resulted in larger numbers of regions with high particle counts (as demonstrated in Figure~\ref{fig:exp}(b) and further illustrated in the Supplementary Information), such that the sub-image size used was reduced to $35\times35$ pixels. The PSF width was estimated empirically from measured images using Gaussian fitting on clearly identifiable particle peaks and was found to be $\approx 1.88$~pixels. 

Ideally, nanoparticles (or clusters) deposited onto a surface would be distributed with uniform spatial probability across the imaging field of view. Within any finite region of interest (e.g. a sub-image) the discrete count of particles would hence obey Poisson statistics (for which the variance and mean are equal). Deviations from this ideal dispersion are however commonly induced by several physical processes, including sample tilt, thermal gradients, image distortion, localised depletion of free nanoparticles, coincidence loss, and surface heterogeneity driving preferential binding. Such compounding factors can necessitate the use of the generalized Poisson distribution (GPD) \cite{Consul01111973} to describe count statistics, as defined by its probability mass function
\begin{align}
p_N(N;\meanN, \psi) = \frac{\meanN (\meanN + N \psi)^{N-1} }{N!} \exp[-\meanN - N \psi],\label{eq:GPD}
\end{align}
where $\meanN >0$ is the rate parameter (analogous to that of a Poisson process, albeit note that the mean of the GPD is $\meanN / (1-\psi)$), $N$ is the  count and $\psi$ is the so-called dispersion parameter. Over-dispersion, that is to say when $\psi > 0$, whereby the variance of particle count is greater than the mean, can be attributed to, for example, image distortion or surface heterogeneity, whereas under-dispersion ($\psi < 0$, $\text{variance} < \text{mean}$) can be caused by coincidence loss suppressing high-count events when particles strongly overlap in an image. 
To test our algorithm in the context of a diagnostic assay and determine spot density $\meanN$, we perform a non-linear least squares fit of the observed relative frequency of particle counts across each experimental image set (Section~\ref{sec:experimental_method}) using the GPD (Eq.~\eqref{eq:GPD}) comparing nanoparticle-only (control) samples against those with the target coronavirus biomarkers (E-DNA and N-DNA). The fitting results are shown in Figure~\ref{fig:exp}(c)-(e). The error bands shown correspond to the observed standard deviation in relative frequency across the set of obtained images. Note that in the fitting, the $N=5$ relative frequency was excluded since at higher nanoparticle concentrations this hypothesis is also likely selected when $N>5$ particles are present in any given sub-image, which would otherwise skew the distribution.  Excellent agreement between the observed counts and GPD fits are evident in Figure~\ref{fig:exp} with the lowest $R^2$ of 0.985 corresponding to the 13.2~pM total concentration control data. A summary of the fitting parameters found and a comparison to the conventional Poisson distribution fit is given in the Supplementary Information.

While the GPD provides a complete description of the count statistics, this level of full number resolution is often not employed in conventional digital assays. Instead, readout in such platforms is typically simplified to a binary ``on/off'' signal to mitigate the impact of signal noise or instrument limitations that make reliably distinguishing $N=1$ from $N=2$ challenging. Although multi-threshold approaches have been proposed \cite{Jacky2021}, a single signal intensity threshold is drawn to binarize all partitions into  `negative' (corresponding to $N=0$) and `positive' (corresponding to $N \ge 1$) states. Concentrations are then subsequently estimated using the observed null count $\nu_N(0)$, as ${\meanN} = -\log \nu_N(0)$. This is usually argued based on the underlying assumption of Poisson statistics, but we here note that such null count estimators are also robust to under and over-dispersion effects since Eq.~\eqref{eq:GPD} reduces to $p_N(0;\meanN,\psi) = \exp[-\meanN]$ (in a similar fashion to the conventional Poisson distribution), which is independent of the dispersion parameter $\psi$.  A comparison of the extracted rate parameter with the null count method is also given in the Supporting Material.  Close agreement is observed between the GPD and null count $\meanN$ estimates, confirming the robustness of the null count method and the accuracy of our counting algorithm. Conventional Poisson fitting, however, whilst producing similar values of $\meanN$, gave a statistically poor fit to observed counts, as evidenced by the lower $R^2$ values (see Supporting Information), thereby demonstrating the distribution's inability to adequately describe our data.

Figure~\ref{fig:exp}(f) shows the variation of the extracted density parameters $\meanN$ (from the GPD fitting) against the nominal total nanoparticle concentration, exhibiting the expected linear dependence. Note that the $\meanN$ found for the high concentration case (with smaller image regions) was scaled by a factor of $(50/35)^2$ to enable meaningful comparison. Importantly, $\meanN$ represents the  surface density of the deposited nanoparticles (and/or clusters) in a $50\times 50$~pixel region, which is distinct from the bulk concentration in solution. A direct, quantitative conversion between these two measures would in principle require either a standard calibration curve or a mass transport model. Absolute quantification of the nanoparticle reporter is, however, not the objective for this diagnostic application. The goal is instead to determine if the presence of the target analyte induces a statistically robust change in the observable particle deposition statistics. 
The statistical significance of the difference between the control and positive samples was therefore evaluated using the likelihood ratio (G-) test and closely related Chi-squared ($\chi^2$) test. Across all three concentration regimes, the tests yielded p-values effectively equal to zero, thereby providing strong support to reject the null hypothesis that the observed distributions for the control and positive samples are the same, i.e., the target coronavirus DNA induces a statistically reliable change in the observed spot count statistics.

Further insight can be gained from our data by inspection of the fitted spot density $\meanN$ and dispersion parameter $\psi$. In particular we observe a number of trends. Firstly, we note that the spot density for the control sample is lower than that found for the positive case at lower nanoparticle concentrations (3.3~pM and 6.6~pM), yet the converse holds at high concentrations (13.2~pM). This result is counter-intuitive, as the target DNA is expected to induce specific aggregation, which would naively suggest a decrease in spot density relative to the control for all cases. We attribute this trend to the complex competition between specific (target-induced) and non-specific aggregation pathways. In spite of efforts to reduce non-specific particle binding (see Supplementary Information), this phenomenon, driven by van der Waals forces and/or salt-induced surface charge screening, nevertheless occurs. When the target (strongly negatively charged) DNA is present, whilst inducing some particle clustering, it additionally coats the surface of unlinked particles, hence adding significant charge which acts as a stabilizing agent. This passivation layer protects the nanoparticles from non-specific binding, such that the net clustering is reduced and $\meanN$ increases. At higher nanoparticle concentrations, however, by virtue of greater particle proximity, higher target-analyte binding affinities and potential localized depletion of free nanoparticles, specific binding starts to dominate over non-specific aggregation pathways, such that $\meanN$ drops in the presence of the analyte. Resulting differences in the average aggregate size, and corresponding differences in mass transport may also contribute to the differing linear gradients observed in Figure~\ref{fig:exp}(f).

Secondly, for both the control and target-positive experiments, we consistently observe $\psi > 0$, indicating significant over-dispersion, which moreover increases with nanoparticle concentration. The behaviour of $\psi$ is also seen to switch at higher concentrations (akin to $\meanN$), with larger over-dispersion found for the control case at lower concentrations. The observed over-dispersion is posited to result from a number of factors. 
A concentration independent contribution to $\psi$ likely originates from PSF elongation and vignetting observed in the peripheries of our images. Although images were cropped to mitigate this issue, some residual contribution to spatially variant count statistics may remain. A concentration dependent component to $\psi$ could additionally arise from nanoparticles (and clusters), which have settled on the imaging substrate, serving as seeds, i.e. preferential binding sites, for further aggregation. The relative importance of this effect would be dependent on the dominant aggregation pathway (specific vs. non-specific) as discussed above. Differential rates of cluster sedimentation onto the imaging plane may also be a contributing factor. Note that given the observed particle densities $\meanN$, coincidence loss (which would reduce $\psi$) is expected to be minimal.

\section{Conclusion}

In this work, we have developed a hypothesis-based particle detection algorithm that provides accurate and interpretable particle counts without reliance on training data or empirical thresholds. By modeling pixel intensities as Poisson random variables tied to a Gaussian point spread function, the framework links inference directly to physical imaging parameters and statistical decision theory. Although specific imaging assumptions were made reflecting typical flourescence and nanoparticle imaging modalities, it should be observed that the approach is highly generalizable to account for engineered PSFs, distinct imaging modalities or noise regimes. 

Systematic performance simulations were presented, which enabled us to benchmark and define the algorithm's optimal operating range. We demonstrated robust accuracy under weak signals, elevated backgrounds, changes in magnification, and moderate PSF mismatch. Two-particle resolution tests highlighted predictable error modes, with under-counting dominating at very small separations and localized over-counting near the resolvability threshold. These results establish clear boundaries for resolvability and clarify the algorithm’s conservative failure modes.

Finally, application of the proposed algorithm to experimental dark-field images served as a practical demonstration. Particle counts were shown to follow  expected statistical distributions and, critically, statistically significant count statistics were observed for samples in which target coronavirus derived DNA biomarkers was present. Count statistics obtained moreover provided insight into competition between specific and non-specific particle binding in our imaging assays, knowledge of which can be leveraged for assay optimization.

Together, these numerical and experimental findings establish our method as a reliable tool for quantitative nanoparticle imaging assays where statistical fidelity is essential. Future work will extend the domain of application to color images, such that spectral information can be leveraged for instance for multiplexed diagnostics, and to include more advance imaging modalities, such as interferometric scattering microscopy. 


\vfill

\end{document}


\title{Supplementary Information for\\
Hypothesis-Based Particle Detection for Accurate Nanoparticle Counting and Digital Diagnostics}

\author{Neil~H.~Kim\orcidlink{0000-0001-9966-4998}, Xiao-Liu Chu\orcidlink{0000-0002-7476-2097},   Joseph B. DeGrandchamp\orcidlink{0000-0003-2194-5495}, Matthew~R. Foreman\orcidlink{0000-0001-5864-9636}
 \thanks{The authors are with the Institute for Digital Molecular Analytics and Science, Nanyang Technological University, 59 Nanyang Drive, Singapore 636921, Singapore. M.R. Foreman is also with the School of Electrical and Electronic Engineering Department, Nanyang Technological University, 50 Nanyang Avenue, Singapore 639798. email: matthew.foreman@ntu.edu.sg}
 }
     
\maketitle

\section{Peak Pixel Fraction under Gaussian PSF}

For a Gaussian point spread function (PSF) of width $\sigma$, the fraction of total signal intensity captured by a single pixel depends on the relative alignment of the PSF center and the pixel grid. The pixel-integrated fraction for a pixel centered at the origin with the PSF centered at offset $(\Delta x,\Delta y)$ is
\begin{align}
f(\sigma;\Delta x,\Delta y)=f_x(\sigma;\Delta x)f_y(\sigma;\Delta y)
\end{align}
where 
\begin{align}
f_s(\sigma;\Delta s) = \frac{1}{\sqrt{2\pi\sigma^{2}}}
  \int_{\substack{-0.5}}^{\substack{0.5}} \exp\!\left[-\frac{(s-\Delta s)^{2}}{2\sigma^{2}}\right]\,ds.
\end{align}
For $\sigma = 2$ pixels and constraining the PSF centroid to lie within the domain of the pixel under consideration, numerical evaluation gives
\[
\begin{aligned}
f_{\min} &\approx 0.0366 \ (3.66\%),\\
f_{\text{avg}} &\approx 0.0378 \ (3.78\%),\\
f_{\max} &\approx 0.0390 \ (3.90\%).
\end{aligned}
\]
Assuming a particle scattering strength of $I = 20{,}000$, the corresponding expected pixel counts are
\[
\begin{aligned}
I  f_{\min} &\approx 732,\\
I  f_{\text{avg}} &\approx 756,\\
I  f_{\max} &\approx 780.
\end{aligned}
\]
At the baseline background level $\lambda_{BG} = 2{,}000$, the peak-to-background ratios are hence
\[
\begin{aligned}
(I f_{\min})/\lambda_{BG} &\approx 0.366,\\
(I f_{\text{avg}})/\lambda_{BG} &\approx 0.378,\\
(I f_{\max})/\lambda_{BG} &\approx 0.390.
\end{aligned}
\]

\section{Jacobian and Hessian}
As described in the main text, we optimize the penalized negative log-likelihood
\begin{equation}
l_p(\boldsymbol{\theta})=\sum_{i,j}\!\bigl[\lambda_{ij}-v_{ij}\log \lambda_{ij}\bigr]+
\alpha\mathcal P(\boldsymbol{\theta}),
\end{equation}
with
\begin{equation}
\lambda_{ij}=\lambda_{BG}+\sum_{k=1}^{N} I_k\,b_k(i)\,a_k(j),
\end{equation}
where $a_k(j) = f_x(\sigma;\Delta x_k^p)$ and $b_k(i)= f_y(\sigma;\Delta y_k^p)$ are the pixel-integrated 1D Gaussian PSF factors in $x$ and $y$ for the $k$-th particle with position $(\Delta x^p_k,\Delta y_k^p)$ measured from the center of the pixel. 
The corresponding Jacobian vector and Hessian matrices for the penalized negative log-likelihood are provided to the solver and are explicitly given below.

The Jacobian vector of $l_p$ taken with respect to the parameter vector $\boldsymbol{\theta}$ is defined as
\begingroup
\setlength{\arraycolsep}{3pt}
\begin{equation}
\resizebox{.82\linewidth}{!}{$
\mathbf J=\frac{\partial l_p}{\partial \boldsymbol{\theta}}=
\begin{bmatrix}
\frac{\partial l_p}{\partial \lambda_{BG}} &
\frac{\partial l_p}{\partial I_{1}} &
\frac{\partial l_p}{\partial x_{1}^p} &
\frac{\partial l_p}{\partial y_{1}^p} &
\cdots &
\frac{\partial l_p}{\partial I_{n}} &
\frac{\partial l_p}{\partial x_{n}^p} &
\frac{\partial l_p}{\partial y_{n}^p}
\end{bmatrix}
,$}
\end{equation}
\endgroup
where $(x_k^p, y_k^p)$ is the global position of the $k$th particle. The partial derivatives can be expressed in the form
\begin{align}
\frac{\partial l_p}{\partial \lambda_{BG}}&=\sum_{i,j} q_{ij},\\
\frac{\partial l_p}{\partial I_k}&=\sum_{i,j} q_{ij}\,b_k(i)\,a_k(j),\\
\frac{\partial l_p}{\partial x_k^p}&=\sum_{i,j} q_{ij}\,I_k\,b_k(i)\,a_k'(j),\\
\frac{\partial l_p}{\partial y_k^p}&=\sum_{i,j} q_{ij}\,I_k\,b_k'(i)\,a_k(j),
\end{align}
where for convenience we define 
\begin{equation}
a_k'(j)=\frac{\partial f_x(j;\Delta x_k^p)}{\partial x_k^p},\qquad b_k'(i)=\frac{\partial f_y(i;\Delta y_k^p)}{\partial y_k^p}
\end{equation}
and
\begin{equation}
q_{ij}=1-\frac{v_{ij}}{\lambda_{ij}},\qquad r_{ij}=\frac{v_{ij}}{\lambda_{ij}^{2}}.
\end{equation}

\begin{figure*}[ht]
  \begin{center}
  \includegraphics[width=\textwidth]{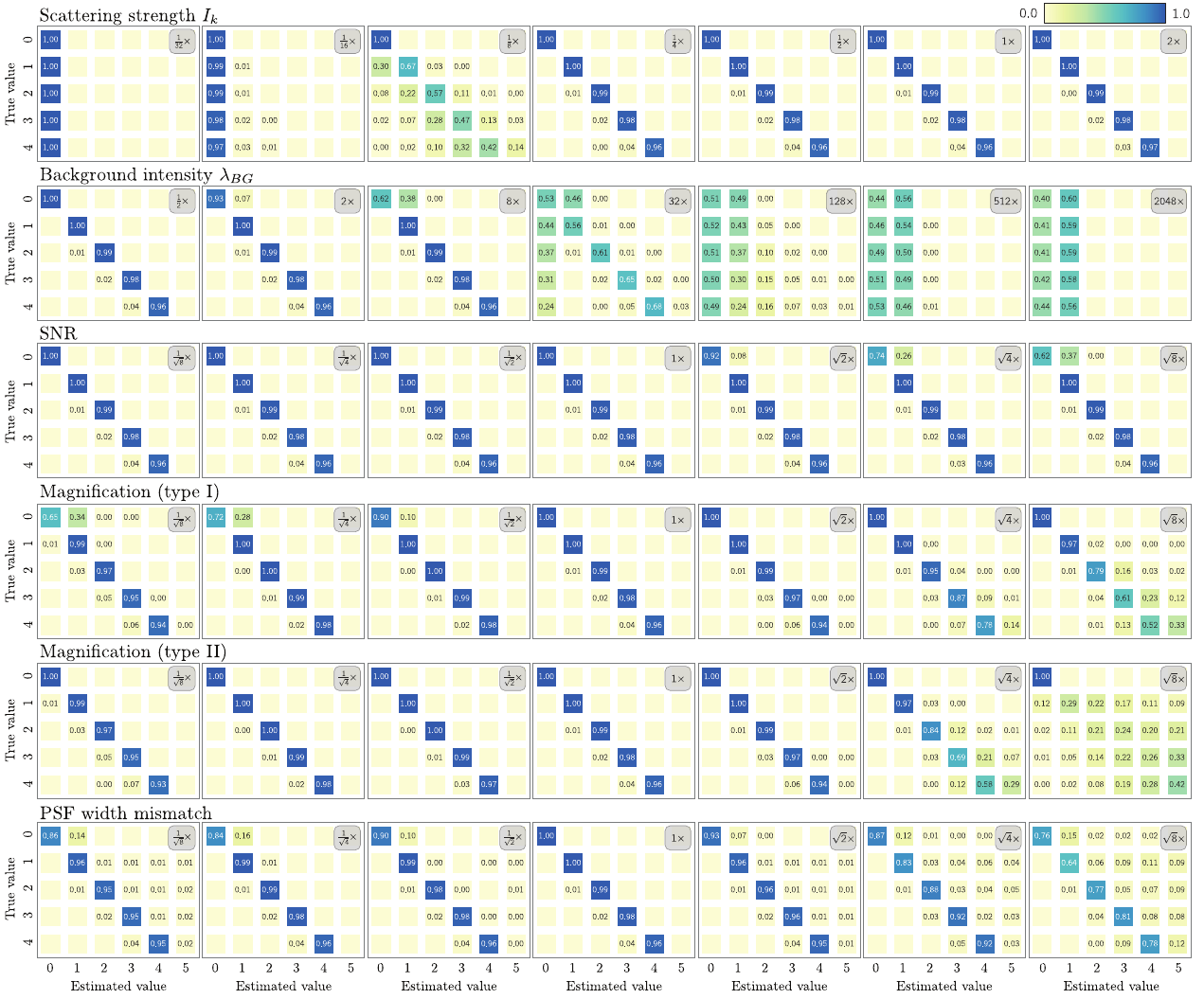}
\caption{\textbf{Simulated confusion matrices.} Representative confusion matrices compiled from the simulated experiments. Each confusion matrix lists true particle count (rows) versus estimated particle count (columns), aggregated over 10{,}000 images per condition. }
\label{sfig:confusion}
  \end{center}
\end{figure*}

Similarly, the Hessian matrix of $l_p$ taken with respect to the parameter vector $\boldsymbol{\theta}$ is
\begingroup
\setlength{\arraycolsep}{3pt}
\begin{equation}
\resizebox{.82\linewidth}{!}{$
\mathbb{H}=\frac{\partial^{2}l_p}{\partial \boldsymbol{\theta}^{2}}=
\begin{bmatrix}
\frac{\partial^{2}l_p}{\partial \lambda_{BG}^{2}} &
\frac{\partial^{2}l_p}{\partial \lambda_{BG}\partial I_{1}} &
\frac{\partial^{2}l_p}{\partial \lambda_{BG}\partial x_{1}^p} &
\cdots \\
\frac{\partial^{2}l_p}{\partial I_{1}\partial \lambda_{BG}} &
\frac{\partial^{2}l_p}{\partial I_{1}^{2}} &
\frac{\partial^{2}l_p}{\partial I_{1}\partial x_{1}^p} &
\cdots \\
\vdots & \vdots & \vdots & \ddots
\end{bmatrix},
$}
\end{equation}
\endgroup
for which representative nonzero blocks take the form
\begin{align}
\frac{\partial^{2}l_p}{\partial \lambda_{BG}^{2}}&=\sum_{i,j} r_{ij},\\
\frac{\partial^{2}l_p}{\partial I_k^{2}}&=\sum_{i,j} r_{ij}\,[b_k(i)a_k(j)]^{2},\\
\frac{\partial^{2}l_p}{\partial I_k\,\partial \lambda_{BG}}&=\sum_{i,j} r_{ij}\,b_k(i)a_k(j),
\end{align}
\vspace{-12pt}
\begin{align}
&\frac{\partial^{2}l_p}{\partial x_k^{p2}}
= \sum_{i,j}\Bigl[r_{ij}\,(I_k b_k(i)a_k'(j))^{2}
+ q_{ij}\,I_k\,b_k(i)\,a_k''(j)\Bigr], \label{eq:hxx}\\
&\frac{\partial^{2}l_p}{\partial y_k^{p2}}
= \sum_{i,j}\Bigl[r_{ij}\,(I_k b_k'(i)a_k(j))^{2}
+ q_{ij}\,I_k\,b_k''(i)\,a_k(j)\Bigr], \label{eq:hyy}\\
&\frac{\partial^{2}l_p}{\partial y_k^p\,\partial x_k^p}
= \sum_{i,j} r_{ij}\,I_k^{2}\,b_k'(i)\,a_k'(j)\,b_k(i)\,a_k(j). \label{eq:hxy}
\end{align}
where 
\begin{equation}
a_k''(j)=\frac{\partial^{2}f_x(j;\Delta x_k^p)}{\partial x_k^{p2}}, \qquad
b_k''(i)=\frac{\partial^{2}f_y(i;\Delta y_k^p)}{\partial y_k^{p2}}.
\end{equation}
Other entries follow by symmetry. Notably, cross-particle blocks are zero since $\lambda_{ij}$ is additive with respect to particle index.

The derivatives given above are expressed in terms of true physical parameters (position, scattering intensity etc). In our practical implementation of the algorithm, additional parameter normalization is used to aid convergence. This normalization introduces additional scale factors for the background, intensity, and particle positions. The penalty term contributes additional derivative terms $\alpha\nabla\mathcal P$ and $\alpha\nabla^{2}\mathcal P$ (which are trivial to evaluate) only when the particle positions fall outside the region of interest. 

\section{Confusion Tables}

Confusion tables for the accuracy tests described in the main text are shown in Figure~\ref{sfig:confusion}.
As discussed in the main text at very high background levels (128$\times$ and above), the confusion matrices show that nearly all predictions fall into the $\hat{N}=0$ and $\hat{N}=1$ columns, with the relative ratio between 0 and 1 fluctuating across background levels. Predictions with $\hat{N}\geq2$ are much less frequent, typically with relative frequency on the order of $10^{-3}$ at 2048$\times$ background. This indicates that in the extreme high-background regime the estimator functions effectively as a binary classifier, toggling between background-only and single-particle fits. 
The rare $\hat{N}=2$ estimates are consistent with the requirement for multiple PSF-like fluctuations within a single image, which occur with far lower probability than a single fluctuation. Shallow local minima in the likelihood surface, or implementation-specific regularization may also contribute to these rare events. Although very infrequent, these cases show that the estimator does not collapse exclusively to 0 or 1 predictions.

\section{Calculation of Overlap Probabilities}

\begin{figure}[b!]
  \begin{center}
  \includegraphics[width=1\columnwidth]{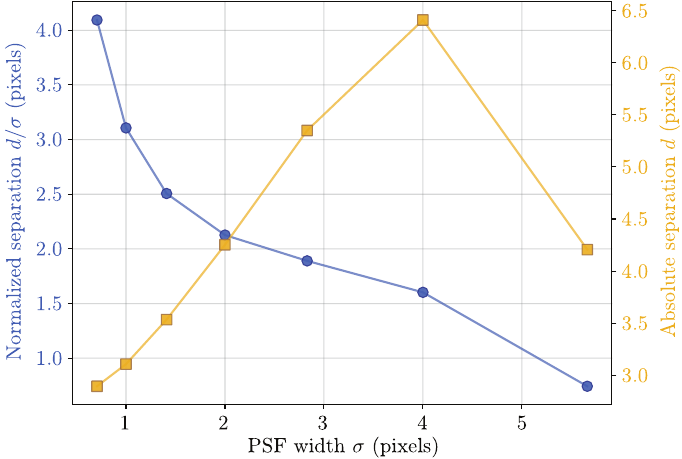}
\caption{\textbf{Resolvability threshold.} Variation of the normalized and unnormalized particle separations at the resolution limit (defined as the separation at which $\hat{N}=2$ estimates fell to 50\% of the total).}
\label{sfig:threshold}
  \end{center}
\end{figure}
To estimate how often close particle pairs occur at $\meanN=1$ in a $100\times100$~pixel image, 
we model particle locations as a homogeneous Poisson point process with rate 
$\rho = \meanN/A = 10^{-4}\,\mathrm{px}^{-2}$ where $A = 10^{4}\,\mathrm{px}^{2}$. The expected number of unordered pairs of particles separated by less than a radius $r$ is
\begin{equation}
\mu_{\mathrm{pairs}} \approx \frac{1}{2}\,\rho^2 A \,\pi r^2.
\end{equation}
The probability of at least one such pair appearing in an image is then
\begin{equation}
p_{\mathrm{image}} \approx 1 - e^{-\mu_{\mathrm{pairs}}} \;\approx\; \mu_{\mathrm{pairs}},
\end{equation}
where the second approximation is valid for the small values of $\mu_{\text{pairs}}$ considered here. Noting that overlap requires $N \ge 2$ particles, we normalize by $P(N\ge2) = 1 - 2e^{-1} \approx 0.264$ to obtain the conditional probability
\begin{equation}
p_{\mathrm{image}\mid N\ge2} \approx \frac{p_{\mathrm{image}}}{P(N\ge2)}.
\end{equation}
For the baseline case $\sigma = 2$~px, Figure 4 of the main text shows that two particles are essentially always resolved once their separation exceeds $r \approx 2.1\sigma$ (see also Supplementary Figure~\ref{sfig:threshold}). 
With $r = 4.2$~px, this gives $\mu_{\mathrm{pairs}} \approx p_{\mathrm{image}} \approx 0.00277$, 
 and 
\[
p_{\mathrm{image}\mid N\ge2} \approx \frac{0.00277}{0.264} \approx 0.0105 \quad (1.05\%).
\]
The unconditional value $p_{\mathrm{image}} \approx 0.00277$ (0.277\%) represents the overall chance of an overlap in a randomly drawn image, including those with $N=0$ or $1$ where overlaps are impossible. The conditional value $p_{\mathrm{image}\mid N\ge2} \approx 0.0105$ (1.05\%) instead quantifies the probability of overlap given that at least two particles are present, and is the more relevant measure when assessing resolvability in multi-particle images.

\section{DNA Functionalization of Gold and Silver Nanoparticles}

\begin{table*}[h!]
\centering
\caption{ssDNA sequences for Target-DNA and Probes \label{stab:DNA}}
\label{tab:DNAseq}
\begin{tabular}{|c|p{7.5cm}|}
\hline &\\[-6px]
\textbf{ssDNA Strand ID} & \textbf{Sequence (5' to 3')} \\[2px]
\hline\hline &\\[-6px]
E-DNA  & agagacaggtacgttaatagttaatagcgtacttctttttcttgctttcgtggtattctt \\
(from SARS-CoV-2 \textit{E} Gene; & gctagttacactagccatccttactgcgcttcgattgtgtgcgtactgctgcaatattgtt \\   NCBI \texttt{NC\_045512.2})   & aacgtgagtcttgtaaaaccttctttttacgtttactctcgtgttaaaaatctgaattctt \\
&ctagag \\[2px]
\hline&\\[-6px]
N-DNA & actcaacatggcaaggaagaccttaaattccctcgaggacaaggcgttccaattaacacc \\ (from SARS-CoV-2 \textit{N} Gene;  & aatagcagtccagatgaccaaattggctactaccgaagagctaccagacgaattcgtggtg \\ NCBI \texttt{NC\_045512.2}) & gtgacggtaaaatgaaagatctcagtccaagatggtatttctactacctaggaactgggcc \\ & agaagctggacttcccta \\[2px]
\hline&\\[-6px]
PrE1 & SH-TTTTTTTTTT-aactattaacgtacctgtctct \\
PrE2 & SH-TTTTTTTTTT-aggatggctagtgtaactag\\
PrE3 & SH-TTTTTTTTTT-ctctagaagaattcagattt \\
PrN1 & SH-TTTTTTTTTT-aatttaaggtcttccttgcc \\
PrN2 & SH-TTTTTTTTTT-gtagccaatttggtcatctg \\
PrN3 & SH-TTTTTTTTTT-tagggaagtccagcttctgg \\[2px]
\hline
\end{tabular}
\end{table*}

Spherical gold nanoparticles (AuNPs) (diameter, 50~nm), spherical silver nanoparticles (AgNPs) (diameter, 50 nm), and gold-shell silica-core nanoparticles (AuSiNPs) (80~nm Si core, 20~nm Au shell) each with 40~kDa polyvinylpyrrolidone (PVP) as capping agent were purchased from nanoComposix (USA). NPs were functionalized with thiolated DNA probes (Table \ref{stab:DNA}) (Integrated DNA Technologies, USA) using a modified PVP-assisted conjugation protocol \cite{heo_ultrastable-stealth_2015,xu_rapid_2011}. Probe DNA sequences were chosen semi-empirically for particle stability, with initial designs created using NUPACK Web Application \cite{zadeh_nupack_2011,mark_e_fornace_nupack_2022} and with subsequent testing at high salt after conjugation. Thiolated DNA was reduced, for use with AgNPs only, by incubation with tris(2-carboxyethyl)phosphine hydrochloride (TCEP) (Sigma Aldrich) at a 1:100 DNA:TCEP ratio for 1 hr at room temperature. DNA for use with gold was not treated, as TCEP reduction was found to be unnecessary \cite{wu_effects_2019}.  

For conjugation, PVP-capped NPs were first aliquoted from high concentration stocks (AuNP: 27.9 OD @ 528~nm, \textit{est.} 1.2 nM; AgNP: 134.9 OD @ 419 nm, \textit{est.} 2.7 nM; and Au@SiNP: 37.8 OD @ 666 nm, \textit{est.} 0.15~nM) into DNA Lobind tubes (Eppendorf SE, Germany). The respective thiolated probes were added at a final concentration of 28.5 \textmu M followed by bath sonication of the reaction (10~s) \cite{hurst_maximizing_2006} and incubation (10 min) at room temperature. Additional PVP (40 kDa) (Sigma Aldrich) was added to a final concentration of 0.35\% (w/v) to further stabilize the particles. A volume of salt-containing buffer, consisting of 20 mM sodium phosphate buffer (PB, pH 8), 200 mM NaBr, and 0. 02\% (w/v) sodium dodecyl sulfate (SDS) (Sigma Aldrich), equivalent to the reaction volume was slowly added in a single step. Notably, NaBr was used instead of NaCl to discourage nonspecific adhesion of nucleotides to the gold/silver surface \cite{liu_bromide_2018}. The reaction mixture was then heated at 60 \textdegree C for 2.5 hr with mixing (800 rpm). To remove residual DNA and salts, the resulting conjugated NPs were washed 5 times (or 7 times for viral detection) by centrifugation (18000\textit{g}, 5 min) using 10 mM PB (no NaCl). Conjugated NP concentrations were then approximated using their UV-Vis Absorbance spectra (NanoDrop One Instrument, Thermofisher Scientific, USA) as compared to their reported stock concentrations.

\begin{table*}[ht]
\centering
\caption{Summary of fitted parameters for null count, standard Poisson, and GPD models. `---' indicates a parameter is not applicable to that model. For the 13.2 pM case rescaled $\meanN$ are shown in brackets.}
\label{tab:gpd_summary}
\begin{tabular}{|c|c| c | c c c |c c c|}
\hline
&&&&&&&& \\[-2ex]
\textbf{Concentration (pM)} & \textbf{Method} & \textbf{Image size (pixels)} &{$\meanN_\text{control}$} & {$\psi_\text{control}$} & {$R_\text{control}^2$} & {$\meanN_\text{covid}$} & {$\psi_\text{covid}$} & {$R_\text{covid}^2$} \\[1ex]
\hline\hline 
&&&&&&&& \\[-2ex]
3.3                & Null count &$50\times 50$ & 0.335 & {---} & 0.989  & 0.506 & {---} & 0.986  \\[1ex]
6.6                & Null count &$50\times 50$& 0.772 & {---} & 0.923  & 1.185 & {---} & 0.775  \\[1ex]
13.2           & Null count &$35\times 35$ ($50\times 50$) & 1.073 (2.191)  & {---} & 0.442  & 0.900 (1.837)& {---} & 0.617  \\[1ex]
\hline
&&&&&&&& \\[-2ex]
3.3                & Poisson  &$50\times 50$   & 0.304 & {---} & 0.991  & 0.478 & {---} & 0.986  \\[1ex]
6.6                & Poisson  &$50\times 50$   & 0.742 & {---} & 0.920  & 1.229 & {---} & 0.787  \\[1ex]
13.2          & Poisson &$35\times 35$ ($50\times 50$)    & 1.092 (2.228) & {---} & 0.495  & 0.858 (1.751) & {---} & 0.671  \\[1ex]
\hline
&&&&&&&& \\[-2ex]
3.3                & GPD  &$50\times 50$      & 0.334 & 0.270 & 1.000  & 0.501 & 0.193 & 0.999  \\[1ex]
6.6                & GPD  &$50\times 50$      & 0.770 & 0.295 & 0.998  & 1.184 & 0.264 & 0.998  \\[1ex]
13.2           & GPD  &$35\times 35$ ($50\times 50$)       & 1.087 (2.218) & 0.432 & 0.985  & 0.909 (1.854) & 0.489 & 0.993  \\[1ex]
\hline

\end{tabular}
\end{table*}

A 188 bp single-stranded DNA sequence (``E-DNA") derived from the Envelope Gene and 200 bp DNA sequence (``N-DNA") derived from the Nucleocapsid Gene of SARS-CoV-2 (NCBI \texttt{NC\_045512.2}) were designed for use as model DNA targets (see Table \ref{tab:DNAseq}) (Ultramer ssDNA, PAGE purified; Integrated DNA Technologies, USA). We used these two strands at equivalent concentrations to emulate a larger SARS-CoV-2 target and refer to them collectively as the ``Target-DNA." For Target-DNA detection, particles were functionalized with two probes each (PrE1 and PrN1 for AuSiNPs; PrE2 and PrE3 for AgNPs; PrN2 and PrN3 for AuNPs), which all bind to specific Target-DNA sites and enable target-induced crosslinking. The exact sequences are found in Table \ref{tab:DNAseq}. For the assay, the nanoparticle mixture (in PB), a reaction buffer mixture, and Target-DNA (in TE buffer) at the relevant concentrations were prepared. The three separate solutions were mixed to start the hybridization assay. The reaction was set up such that the following final concentrations were achieved in a 40 \textmu L volume: 10 mM PB, 400 mM NaCl, 2 mM Ethylenediaminetetraacetic acid (EDTA) (Sigma Aldrich), 3\% (w/v) dextran (150 kDa) (Sigma Aldrich), 0.05\% (w/v) PVP (40 kDa), 0.05\% (w/v) SDS, 10 pM Target-DNA ([E-DNA] = [N-DNA] = reported concentration) or TE buffer as control. EDTA was included to chelate undesirable multivalent ions, while dextran was included to promote specific hybridization \cite{amasino_accel_1986,donnelly_investigation_2016}. Salt and surfactant concentrations were empirically derived for optimal hybridization results (fast and specific). The nanoparticle concentrations were varied as follows:
\begin{enumerate}
    \item Low: AuSiNPs: 0.3~pM, AuNPs and AgNPs: 1.5~pM each.
    \item Medium: AuSiNPs: 0.6~pM, AuNPs and AgNPs: 3~pM each.
    \item High: AuSiNPs: 1.2~pM, AuNPs and AgNPs: 6~pM each.
\end{enumerate}
Immediately after mixing, the solution was vortexed and then heated to 50 \textdegree C with mixing (800 rpm) for 15 min. The reaction was allowed to cool for 10 min at room temperature before transferring 20 \textmu L to a home-made microscopy chamber consisting of two plasma-etched coverslips sandwiching a parafilm chamber for optical imaging. Notably, single AuSiNPs settle via gravity onto the glass surface, while other particles tend to stay diffuse in solution unless they are clustered. Images were captured in a homebuilt darkfield microscope.

\section{Generalized Poisson distribution fitting}
\begin{figure}[ht]
  \begin{center}
  \includegraphics[width=\columnwidth]{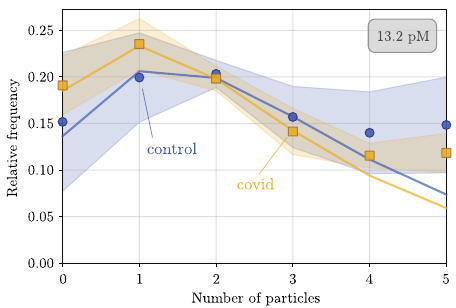}
\caption{\textbf{Nanoparticle imaging based SARS-CoV-2 assay - high nanoparticle concentration case.} Distribution of estimated particle counts for the high nanoparticle concentration case when $50\times50$ sub-images were used, for both control (blue markers) and SARS-CoV-2 positive (orange markers) samples, overlaid with GPD fits (solid curves). Error bands correspond to inter-image standard deviations. Significant counts at $\hat{N}=5$ are evident motivating reduction of the sub-image size.}
\label{sfig:high50}
  \end{center}
\end{figure}

Table \ref{tab:gpd_summary} details the fitted parameters ($\meanN$ and $\psi$ where appropriate) and the resulting $R^2$, for the null count, Poisson, and GPD models across the tested concentration range. For the highest concentration (13.2 pM), as detailed in the main text, the final analysis was performed on reduced $35\times35$ pixel sub-images. This was motivated by the results in shown in Figure \ref{sfig:high50}, which used a $50\times50$ sub-image size, in which significant counts at the maximum hypothesis number $n_{\text{max}} =5$ were seen. Table \ref{tab:gpd_summary} includes  the raw extracted $\meanN$ values and scaled values (in brackets) for this case, so as to facilitate direct comparison.

\vfill